\documentclass[amssymb,prb,aps,twocolumn,showpacs,10pt]{revtex4}
\usepackage[dvips]{graphicx}
\begin{document}
\author{E. Cohen$^1$, S. Muenzel$^2$, J. Fleischer$^2$, V.
Fleurov$^1$, A. Soffer$^3$}
\affiliation{$^1$Raymond and Beverly Sackler Faculty of Exact
Sciences, School of Physics and Astronomy,\\ Tel-Aviv University,
Tel-Aviv 69978 Israel.\\ $^2$ Department of Electrical Engineering,
Princeton University, Princeton, NJ 08544, USA \\$^3$Department of
Mathematics, Rutgers University, New Brunswick, NJ 08903,USA}

\title{Jet-like tunneling from a trapped vortex}

\author{}
\affiliation{$^1$Raymond and Beverly Sackler Faculty of Exact
Sciences, School of Physics and Astronomy,\\ Tel-Aviv University,
Tel-Aviv 69978 Israel.\\ $^2$Department of Mathematics, Rutgers
University, New Brunswick, NJ 08903,USA}

\begin{abstract}

We analyze the tunneling of vortex states from elliptically shaped
traps. Using the hydrodynamic representation of the Gross-Pitaevskii
(Nonlinear Schrцdinger) equation, we derive analytically and
demonstrate numerically a novel type of quantum fluid flow:  a
jet-like singularity formed by the interaction between the vortex
and the nonhomogenous field.  For strongly elongated traps, the
ellipticity overwhelms the circular rotation, resulting in the
ejection of field in narrow, well-defined directions. These jets can
also be understood as a formation of caustics since they correspond
to a convergence of trajectories starting from the top of the
potential barrier and meeting at a certain point on the exit line.
They will appear in any coherent wave system with angular momentum
and non-circular symmetry, such as superfluids, Bose-Einstein
condensates, and light.

\end{abstract}
 \maketitle

Topological charges, such as vortices, are fundamental to the
dynamics of coherent fields\cite{CD99,FBCMS04}. They appear in laser
systems, carry charge in superconductors, characterize turbulence in
quantum fluids, and hold potential for quantum memory\cite{ZSTS07}.
To date, the main focus in vortex dynamics has been on transport, so
that the charges could move and interact. (see e.g. Ref.
\onlinecite{AAG06}) However, it is often desirable, and sometimes
necessary, to confine and trap vortex structures. This is a basic
problem in trapping theory, yet it has received very little
attention. Here, we consider the dynamics of vortex decay in a
potential and show that asymmetry in the potential can lead to the
development of jets during wave tunneling.  These formations
concentrate wave density in the form of caustics and represent a new
type of coherent structure for wave transport.

The emphasis on vorticity implies that phase dynamics will be
important to the tunneling process. Even in the context of simple
wavefunctions, without angular momentum, phase can have profound
effects.  Examples include the recent prediction of "blips" in the
outgoing matter through a trap\cite{DFSS07,bpss08,bpsss08,DFSF09}
and the development of dispersive shock
waves\cite{shock1,shock4,shock5,shock6}, e.g. when tunneling through
a barrier.\cite{DFSF09a} These latter structures are traveling waves
with oscillating phase that are finding increasing importance in
fluids\cite{B04,LYP04}, optics \cite{FBCMS04,shock5}, and
Bose-Einstein condensates\cite{vortex1,vortex4}. In spatially
inhomogeneous potentials, such as the elliptical wells typical of
BEC experiments \cite{GBH05}, both shock waves and blips can go
unstable and generate vortices. Here, we consider the simplest case
of a circular vortex trapped in an elliptical well and examine the
competition of symmetry during wavefunction tunneling.
\begin{figure}[tb!]
\begin{center}
\includegraphics[width=2.5in,keepaspectratio]{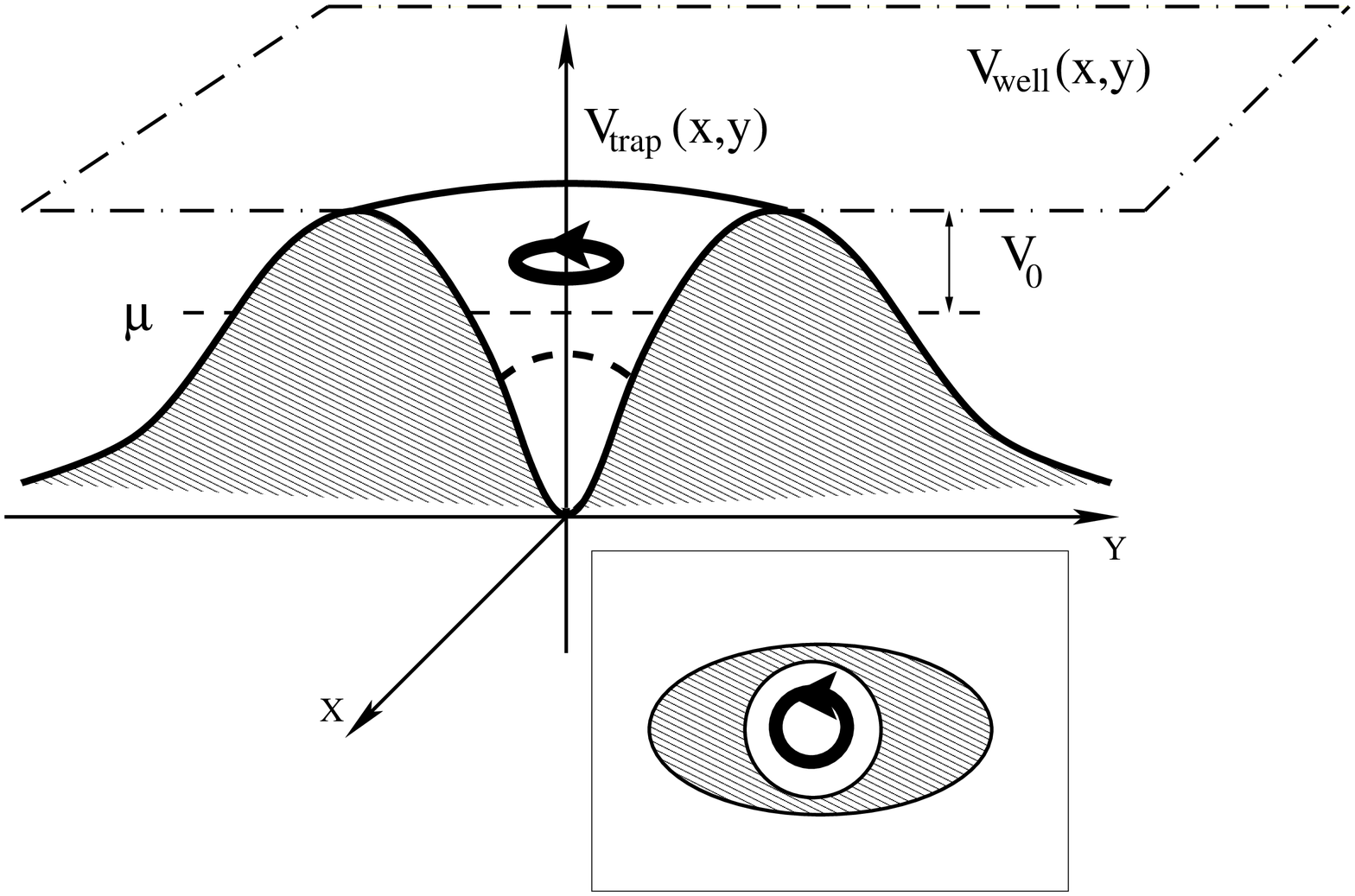}
\end{center}
\caption{A cross-section of the potential $V_{trap}(x,y)$ of an
elliptical trap with cylindrically symmetric internal potential
well, which holds a vortex state shown symbolically. The potential
$V_{well}(x,y)$, which tends to a constant value for large $x$
and/or $y$ (shown by dashed-doted lines) is used for the preparation
of the initial state. $V_0$ is the difference between the top rim of
the barrier and the chemical potential level $\mu$ whose crossing
with the barrier determines the exit line. The insert shows a view
from above with the circular top rim, a vortex inside it, and the
elliptic exit line outside. } \label{pc}
\end{figure}

Tunneling problems are usually discussed within the framework of the
WKB approximation, which looks for a solution of the Schr\"odinger
equation (Nonlinear Schr\"odinger (NLS) or Gross-Pitaevskii (GP)
equation in our case)
\begin{equation}\label{NLS}
i\hbar \frac{\partial \psi}{\partial t} = - \frac{\hbar^2}{2m}
\nabla^2 \psi + V_{trap}({\bf r}) \psi + \lambda |\psi|^2\psi
\end{equation}
as an expansion over $\hbar$. In the case of the NLS equation
describing propagation of a classical coherent electromagnetic wave
$\hbar =1$, $\psi$ is the amplitude of the electric component, the
propagation distance $z$ plays part of time, the part of mass $m$ is
played by the wave vector of the light emitted by the laser, and the
potential $V$ is created by changing the linear refraction index of
the medium. The appearance of singularities near the turning points
is typical for WKB approach. These singularities make generalization
to 2d and higher dimensions a formidable technical problem (see,
e.g. Ref. \onlinecite{I09} and references therein) since then the
points become singular lines or surfaces. (Nevertheless calculations
can be quite straightforward in the cylindrical symmetry
case\cite{chm05} even from the vortex state.) The most promising way
to deal with this problem is to use the hydrodynamic approach, which
was applied to the tunneling dynamics in Refs.
\onlinecite{FS05,DFSS07} [See also Refs.
\onlinecite{shock1,shock4,shock5,shock6}]. Two hydrodynamic
equations
\begin{equation}\label{continuity-1}
\frac{\partial}{\partial t}\rho + \nabla(\rho {\bf v}) =0
\end{equation}
and
\begin{equation}
\frac{\partial}{\partial t} {\bf v} + \frac{1}{2}\nabla {\bf v}^2 =
- \frac{1}{m}\nabla\left(V_{trap}({\bf r}) - \frac{\hbar^2}{2m}
\frac{\nabla^2\sqrt{\rho}}{\sqrt{\rho}} + \lambda \rho \right)
\label{Euler-1}
\end{equation}
follow directly from the GP equation (\ref{NLS}) for the density
$\rho = |\Psi|^2$ and the velocity $m {\bf v} = - \hbar \nabla
\varphi$ fields determined by the phase $\varphi$ of the wave
function $\Psi$. The principal advantage of this approach is the
absence of any singular behavior of the wave function near the
turning points. For example, the temporal tunneling dynamics was
deduced analytically\cite{FS05,DFSS07} from these equations
accounting for the role of the interaction (nonlinearity) in 1d
systems.

Here we consider wave tunneling from a two-dimensional trap.  We
first consider an irrotational initial state in order to present our
approach and for the sake of comparison.  We then generalize it to
the rotational case with a vortex initial state. We will show that
in the case of nonzero angular momentum, jets appear in which
matter/intensity exit the trap along certain preferential
directions.

In the 1d case\cite{FS05} (see also a more detailed derivation in
Ref. \onlinecite{DFSS07}), the adiabatic approximation in the Euler
equation (\ref{Euler-1}) yields the velocity at the exit point in
the form $v(x_{ex}) = \sqrt{2 V_0/m}$, where $V_0$ is the energy
difference between the top of the barrier and the chemical potential
in the trap. The same definition holds for the 2d case (see Fig.
\ref{pc}). In this higher dimension, there is an exit curve where
the chemical potential $\mu$ crosses the trap potential. Integrating
the exit flux $\rho({\bf r}){\bf v}_{ex}({\bf r})\cdot \hat{n}$ over
this closed curve ($\hat{n}$ is the unit vector normal to this
curve), gives the total exit flux
\begin{equation}\label{rate-2}
\frac{dN(t)}{dt} = -\oint\rho({\bf r}) ({\bf v}_{ex}({\bf r})
\cdot\hat{n}) d{\bf l}.
\end{equation}
The 2d tunneling problem\cite{DFSS07} is mapped onto the classical
motion of a fluid droplet (tracer) falling down from the top of the
difference potential $\Delta V(x,y) = V_{trap}(x,y) - V_{well}(x,y)$
to the observation points $(x,y)$ on the exit line. Here,
$V_{well}(x,y)$ is the potential well used to prepare the initial
state and $V_{trap}(x,y)$ is the actual potential of the trap from
which the tunneling takes place (see Fig. \ref{pc}). In the case of
an irrotational flow, the exit velocity vector is found from the
equation $ m \ddot{\bf r} = - {\bf \nabla} \Delta V({\bf r}) $.
Considering the example of the elliptic trap shown in Fig. \ref{pc},
\begin{equation}\label{elliptic}
\Delta V(x,y) = - \frac{m\omega^2}{2}(y^2 + \epsilon^2 x^2)
\end{equation}
with the aspect ratio $\epsilon \leq 1$ we get ${\bf v}_{ex}(x,y) =
(\epsilon \omega x, \omega y)$.

Adiabatically slow varying density\cite{FS05,DFSS07} in the trap may
be described as $ \rho(N,{\bf r},t) = N(t) \rho_0
\exp(-\frac{2}{\hbar} \sqrt{2m V_0}|{\bf r}|)$ and Eq.
(\ref{rate-2}) becomes
\begin{equation}\label{Integ}
\frac{dN}{dt} = - N I_{irrot}
\end{equation}
where $I_{irrot}$ is the integral escape rate. Using the polar
coordinates, $(r,\theta)$, the differential escape rate reads
\begin{equation}\label{escaperate}
\frac{dI_{irrot}}{d\theta} = \frac{2V_0}{m\omega} \rho_0
\frac{\sin^2\theta + \epsilon^3\cos^2\theta}{\chi^2(\varepsilon,
\theta)} e^{-\frac{4\Delta u}{ \sqrt{\chi(\varepsilon, \theta)}}}
\end{equation}
where $\Delta u = V_0/(\hbar\omega)$ and $\chi(\varepsilon, \theta)
= \sin^2\theta + \epsilon^2\cos^2\theta$. This equation is obtained
by using the simple connection $\epsilon \tan \beta = \tan \theta$
between the polar angle $\theta$ and the angle $\beta$ of the exit
velocity direction (see $v_{ex}$ after Eq. 5).

Assuming certain dependence (say, linear) of the interaction on the
number of particles $N$ remaining within the trap, its dependence on
time for various values of the parameters of the system can be found
from Eq. (\ref{Integ}). Qualitatively these dependencies are rather
close to those considered in detail in Ref. [\cite{FS05}],
discussing tunneling from one dimensional traps, although the
numerical values may differ.

A 2d configuration allows the consideration of a rotational initial
state, i.e. one with nonzero vorticity
$\mbox{\boldmath$\widetilde{\omega}$} = \mbox{\boldmath$\nabla$}
\times {\bf v}$,  which is not possible in 1d. Since the velocity
field is $m {\bf v} = - \hbar \mbox{\boldmath$\nabla$} \varphi$, a
finite vorticity in Eq.(\ref{Euler-1}) appears if the phase
$\varphi$ is singular along some lines in 3d space or points in 2d
space. As the wave function is single valued, the equation $ m \oint
{\bf v}({\bf r}) d{\bf l} = 2 \pi \hbar \nu $ with an integer $\nu$
holds for integration over any closed path. The phase $\varphi = -
\frac{2\pi \nu}{m} \arctan\frac{y}{x}$ corresponds to a vortex
around the line $x = y = 0$ with the velocity field $ {\bf
v}_{rot}({\bf r}) = \nu \frac{\hbar}{m \varrho^2} ( - y, x, 0) $
where $\varrho = \sqrt{x^2 + y^2}$. The vorticity reads $
\mbox{\boldmath$\widetilde{\omega}$} = \nu \frac{\hbar}{m } \hat n_z
\delta(x)\delta(y)$. Eq. (\ref{Euler-1}) requires that the density
field vanishes as $\rho(\varrho) \propto \varrho^{2|\nu|}$ at
$\varrho \to 0$ and the quantum potential blows up as
$\varrho^{-2}$.

The total velocity field ${\bf v}({\bf r}) = {\bf v}_{rot}({\bf r})
+ {\bf v}_{pot}({\bf r}) $ is a sum of the rotational and
irrotational velocity fields with $\mbox{\boldmath$\nabla$} \times
{\bf v}_{pot} = 0$. Then we may carry out the same program as above,
i.e. we prepare the initial state in a potential well
$V_{well}(x,y)$ with a vorticity characterized by an integer quantum
number $\nu$. Then we change the potential into $V_{trap}(x,y)$
allowing the wave function to tunnel through the barrier and apply
the adiabatic approximation.\cite{FS05,DFSS07} As a result, the Euler
equation (\ref{Euler-1}) takes the form
\begin{equation}\label{Euler-3}
m \frac{\partial {\bf v}_{pot}({\bf r}, t)}{\partial t} +
\frac{m}{2} \mbox{\boldmath $\nabla$} [{\bf v}_{pot}({\bf r}, t) +
{\bf v}_{rot}({\bf r})]^2 =
$$$$
- \mbox{\boldmath $\nabla$} \left[ \Delta V({\bf r}) -
\frac{\hbar^2\nu^2}{2m} \frac{1}{\varrho^2} \right]
\end{equation}
with the centrifugal potential in the right hand side. Now the Cole
- Hopf transformation\cite{DFSS07} allows one to map the tunneling
problem on the classical motion of a fluid tracer described by the
equation
\begin{equation}\label{eqmotion-2}
m \dot{\bf v}_{rot} + m \dot {\bf v}_{pot} = m {\bf
v}\times\mbox{\boldmath $\widetilde{\omega}$} - \mbox{\boldmath
$\nabla$} \left[ \Delta V({\bf r}) - \frac{\hbar^2\nu^2}{2m
\varrho^2}\right].
\end{equation}
The "Lorentz force" in (\ref{eqmotion-2}) is zero everywhere except
for the line $x=y=0$ and does not play a role.

For ${\bf v}_{pot} = 0$ and $\Delta V({\bf r})= 0 $, one gets $ {\bf
v}_{rot} = \Omega \cdot \hat n_z \times {\bf r}_\Omega(t) $ with $
{\bf r}_{rot}(t) = \varrho \left( \cos\Omega t, \sin\Omega t, 0
\right)$ and $\Omega = \frac{\hbar\nu}{m \varrho^2}$, which
corresponds to  the tracer making a circular rotation with the
velocity $ v = \hbar\nu / m \varrho $. We again use the elliptic
potential (\ref{elliptic}) and consider the simplest vortex with
$\nu = 1$; then the escape rate is calculated similarly to the
rotationless case. We are interested in the angular dependence of
the differential escape rate
\begin{equation}\label{differentialescape}
\frac{dI_{rot}}{d\beta} = \frac{2V_0}{m\omega}  \sqrt{1 +
\frac{4\Delta u^2}{\chi(\varepsilon, \theta)}} \cdot
$$$$
\frac{\epsilon \cos\beta \cos\theta + \sin\beta\sin\theta}{
\sqrt{\chi(\varepsilon, \theta)}}
\left|\frac{d\theta}{d\beta}\right| e^{- \frac{4\Delta
u}{\sqrt{\chi(\varepsilon, \theta)}}}
\end{equation}
%
Contrary to the irrotational case (\ref{escaperate}), there is not
now a simple relation between the polar coordinate $\theta$ and the
angle $\beta$ of the escape direction.  Therefore, we have to use
the more general equation (\ref{differentialescape}) in which the
dependence $\beta(\theta)$ is found by solving equation of motion
(\ref{eqmotion-2}) numerically.

\begin{figure}[ht!]
\begin{center}
\includegraphics[width=2.5in,keepaspectratio]{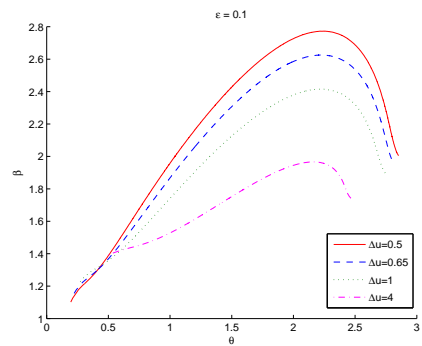}
\end{center}
\caption{Exit angle $\beta$ as a function of polar angle $\theta$
for the barrier hight $\Delta u$ varying from 0.5 to 4 for a highly
eccentric trap (aspect ratio $\epsilon = 0.1$). All the curves pass
maxima indicating the jets coming out of the trap.}
\label{beta(theta)}
\end{figure}

The angular dependence (\ref{differentialescape}) of the escape rate
from a vortex state is the principal result of this paper. It shows
that the rate is determined largely by the function $\beta(\theta)$,
which is not necessarily monotonous. Fig. \ref{beta(theta)} shows
that at small enough aspect ratio $\epsilon$ and moderate barrier
height $V_0$, the escape direction $\beta(\theta)$ obtains a maximum
value $\beta_{max}$ on the exit curve at an angle $\theta_{max}$. At
this point, the derivative $\frac{d\theta}{d\beta}$ blows up,
indicating that an ensemble of streamlines originating from a spread
of $\theta$ angles around $\theta_{max}$ collapses together.  This
bunching results in the formation of a jet in the $\beta_{max}$
direction.

\begin{figure}[ht!]
\begin{center}
\includegraphics[width=3in,keepaspectratio]{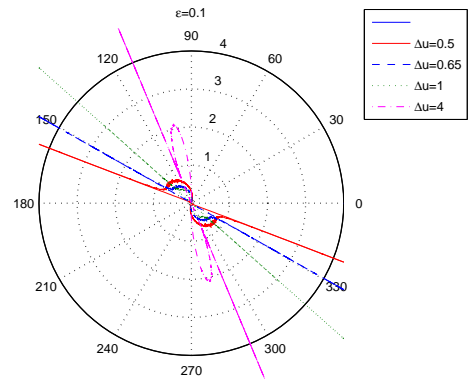}
\end{center}
\caption{Normalized polar graphs of the differential escape rate
$\frac{dI}{d\beta}$ as a function of $\beta$ for $\epsilon = 0.1$.}
\label{figIb,b}
\end{figure}

For small eccentricity and moderate vorticity $\nu$, the exit flow
is distorted only weakly.  For large enough eccentricity and low
barrier heights, or high enough vorticity, this distortion becomes
strong and jets appear. An example of well-developed jets are shown
in Fig. \ref{figIb,b}, obtained when the shape of the trap strongly
deviates from the circular one ($\epsilon = 0.1$).

The jets are caused by the interplay between the elliptical shape of
the trap and the spherical symmetry of the vortex. If the equivalent
tracer motion were dominated by the circular motion only, then we
could have a strange situation in which the tracer could have left
the elongated trap, traveled along its circular orbit, and then
tried to re-enter the trap. Interestingly, this scenario is
prevented by the {\em irrotational} part of the velocity field,
whose contribution leads to a frustration point and the formation of
the jets.  The meeting of many different tracer trajectories in the
vicinity of $\theta_{max}$ on the exit line gives rise to the
formation of caustics.  The resulting interference is similar to the
tunneling dynamics considered in Ref. [\onlinecite{I09}].  As shown
in Fig. \ref{figIb,b}, the variation in tunneling speeds gives rise
to wave steepening and a new type of angular shock.

\begin{figure*}
\hbox{
\includegraphics[width=3in,keepaspectratio]{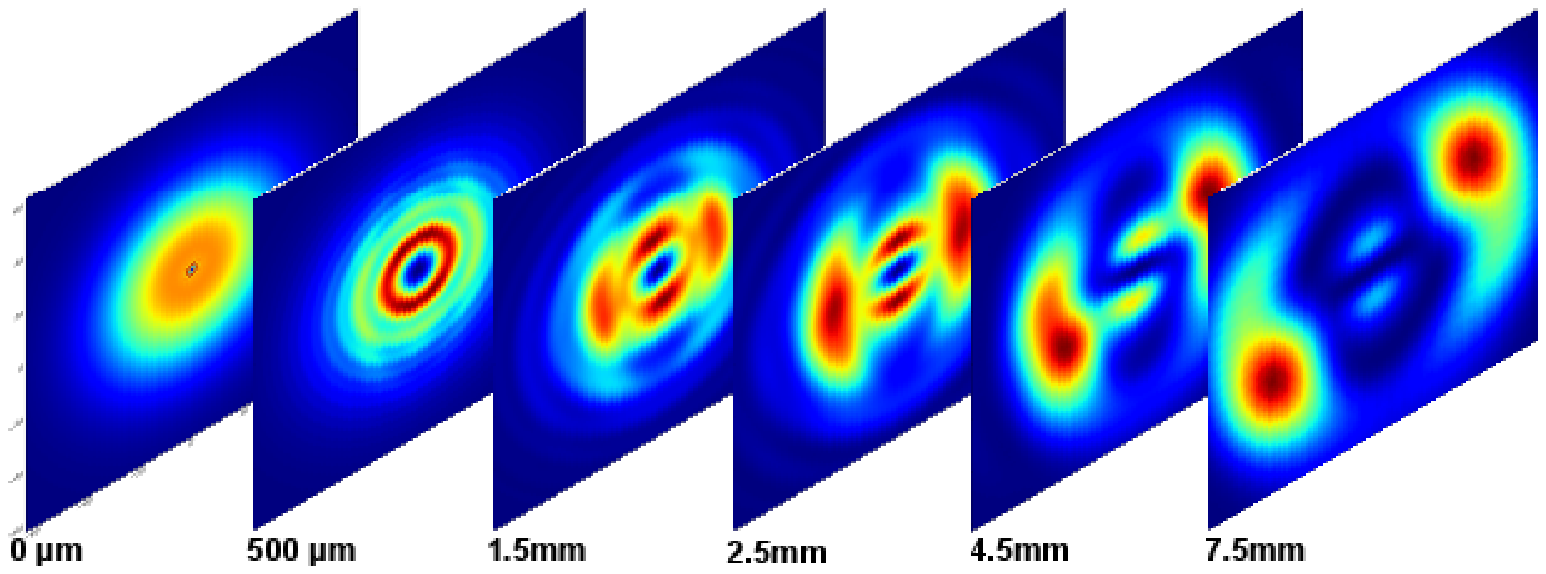} \hspace{2cm}
\includegraphics[width=3in,keepaspectratio]{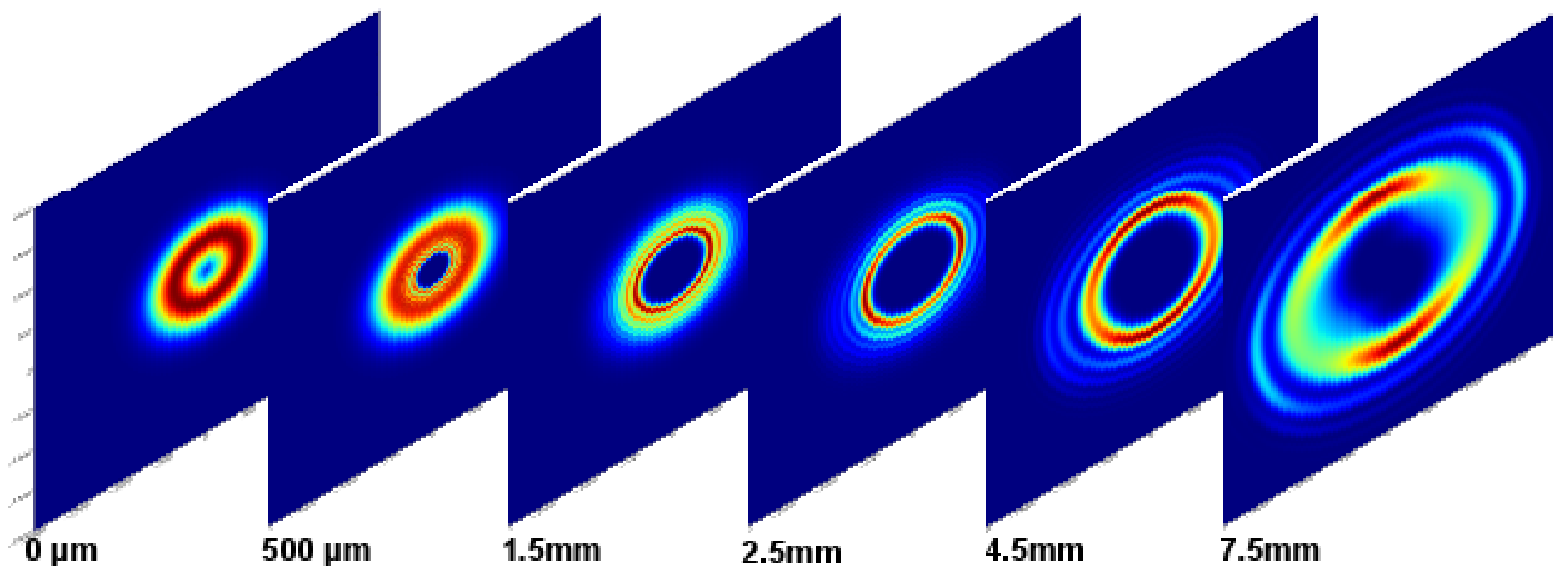}}
\caption{ Numerical calculation of the evolution of a light
propagating in an elliptically shaped semitransparent waveguide for
the vorticity $\nu = 1$ (left) for $\Delta n/n_0 = 5.0\cdot 10^{-5}$ written in PMMA
and $\nu= 8$ for $\Delta n/n_0 = 1.0\cdot 10^{-4}$(right) induced in SBN. Formation
of jets with the propagation of light at a distance 7.5 mm is
clearly seen.} \label{Muenzel}
\end{figure*}

In principle, the exponentially weak tunneling from an elliptically
shaped trap violates the spherical symmetry of the problem and may
cause a decay of the vortex state. In practice, however, the time
span of the decay can be rather long (esp. if the vortex can be
stabilized by a strong enough interaction)\cite{GMPT01}, and
observation of the above dynamics should be possible.  Currently,
the best candidate systems to observe jet-like tunneling are cold
atoms in elliptically trapped Bose-Einstein condensates and coherent
light confined in elongated optical waveguides.

Here, we numerically demonstrate angular caustic formation in the
optical case. The cylindrical waveguide required for an experimental
realization can be written inside a medium such as a glass or
polymethyl methacrylate (PMMA) using femtosecond laser pulses. As a
realistic example, we consider a waveguide with a transverse
refractive index profile (potential well barrier) shown in Fig. 1.
The inner side of the well is circular while the outer edge is an
ellipse with a semi-major axis of 40$\mu$m (corresponding to an
eccentricity $\sqrt{1-\epsilon^2} =0.9$). Simulations were carried
out for the 2+1d system by solving the NLS equation (\ref{NLS})
using a split-step beam propagation code. Results are shown in Fig.
\ref{Muenzel}. Even with a small refractive index difference
(potential barrier height) of $\Delta n/n_0 = 5.0\cdot 10^{-5}$,
jet-like tunneling occurs for a singly-charged vortex $\nu = 1$ (see
Fig. \ref{Muenzel}, left panel). For comparison, we also give
similar results for light propagating in an optically-induced
waveguide in SBN (Strontium Barium Niobate). In this case, the
writing beam diffracts, so the waveguide diameter is limited.  Even
with a 50$\mu$m inner radius, the potential still diffracts.  This
weaker potential means that we cannot see the jet form if the input
vortex is only singly-charged. Shown is the output field of a charge
$\nu = 8$ vortex after propagating 8mm in the crystal (see Fig.
\ref{Muenzel}, right panel).

In summary, by using the hydrodynamic formulation of the nonlinear
Schr\"odinger equation, we were able to carry out an analysis of
tunneling from a vortex field trapped in an asymmetric potential
well. Interference between the rotational motion of the field with a
strongly asymmetric tunneling flow created angular caustics,
resulting in jet-like radiation patterns. Analytic results were
verified with numerical simulation.

{{\bf Acknowledgments.} The authors acknowledge the support of
United States - Israel Binational Science Foundation, Grant N
2006242. A.S. is partially supported by NSF. V.F. and A.S. are
indebted to the hospitality of MPIPKS, Dresden.

\end{document}